\documentclass[aps,prx,twocolumn,superscriptaddress]{revtex4}
\usepackage{amssymb}
\usepackage{amsfonts}
\usepackage{xcolor}
\usepackage{amsmath}
\usepackage{graphicx}
\usepackage{bm}
\begin{document}

\title{Time-resolved investigation of plasmon mode along interface channels

in integer and fractional quantum Hall regimes
}
\author{Chaojing Lin}
\email{lin.c.ad@m.titech.ac.jp}
\affiliation{Department of Physics, Tokyo Institute of Technology, 2-12-1 Ookayama,
Meguro, Tokyo, 152-8551, Japan.}
\affiliation{Tokyo Tech Academy for Super Smart Society, Tokyo Institute of Technology, 2-12-1 Ookayama, Meguro, Tokyo 152-8551, Japan.}
\author{Masayuki Hashisaka}
\affiliation{NTT Basic Research Laboratories, NTT Corporation, 3-1 Morinosato-Wakamiya,
Atsugi, 243-0198, Japan.}
\author{Takafumi Akiho}
\affiliation{NTT Basic Research Laboratories, NTT Corporation, 3-1 Morinosato-Wakamiya,
Atsugi, 243-0198, Japan.}
\author{Koji Muraki}
\affiliation{NTT Basic Research Laboratories, NTT Corporation, 3-1 Morinosato-Wakamiya,
Atsugi, 243-0198, Japan.}
\author{Toshimasa Fujisawa}
\affiliation{Department of Physics, Tokyo Institute of Technology, 2-12-1 Ookayama,
Meguro, Tokyo, 152-8551, Japan.}

\begin{abstract}
Quantum Hall (QH) edge channels appear not only along the edge of the electron gas but also along an interface between two QH regions with different filling factors. However, the fundamental transport characteristics of such interface channels are not well understood, particularly in the high-frequency regime. In this study, we investigate the interface plasmon mode along the edge of a metal gate electrode with ungated and gated QH regions in both integer and fractional QH regimes using a time-resolved measurement scheme. The observed plasmon waveform was delayed and broadened due to the influence of the charge puddles formed around the channel. The charge velocity and diffusion constant of the plasmon mode were evaluated by analyzing the waveform using a distributed circuit model. We found that the conductive puddles in the gated region induce significant dissipation in plasmon transport. For instance, a fractional interface channel with a reasonably fast velocity was obtained by preparing a fractional state in the ungated region and an integer state in the gated region, whereas a channel in the swapped configuration was quite dissipative. This reveals a high-quality interface channel that provides a clean path to transport fractional charges for studying various fractional QH phenomena.
\end{abstract}

%\pacs{73.43.Fj, 73.43.Jn, 73.23.-b, 73.63.-b}
\maketitle

\section{Introduction}

Quantum Hall (QH) edge channels formed along the edge of a two-dimensional electron gas (2DEG) in a high magnetic field govern the transport characteristics of the system \cite{Ezawa13,Halperin82,MacDonald84}. While the linear dc conductance can be explained with a single-particle picture \cite{Landauer70,Buttiker88}, transport in the non-equilibrium and high-frequency regime involves collective excitations called edge magnetoplasmons (chiral plasmons). A charge density wave in the plasmon mode propagates along the channel for a long distance with small damping \cite{Grodnensky91,Ashoori92,Talyanskii94,Talyanskii92,Zhitenev94,Zhitenev95,Ernst96,Ernst97,Kamata10}, indicating that the plasmon approach is appropriate for describing the charge dynamics of the system. Recent experiments have revealed non-trivial many-body effects, such as spin-charge separation \cite{Bocquillon13,Hashisaka17,Itoh18} and charge fractionalization \cite{Kamata14,Inoue14,Lin20}, which can be explained in terms of Tomonaga-Luttinger liquids. The coupling of chiral plasmon modes plays an essential role in these effects. In general, when two QH regions with different Landau-level filling factors are placed side-by-side, a chiral one-dimensional (1D) channel is formed along the interface between them \cite{Lin20,Khaetskii94,Sukhodub04}. This interface channel is essential for studying Tomonaga-Luttinger liquids, as well as hole-conjugate fractional QH states \cite{MacDonald90,Wen90}. Even for a single interface channel, a full understanding of charge dynamics is desirable for transporting fractional charges and heat in a 1D circuit \cite{Lin20,Inoue14nc,Venkatachalam12,Roddaro09}. However, the fundamental transport characteristics of the interface plasmon modes are not well understood, particularly in the high-frequency regime. Unlike edge channels that are confined by a large external confining potential, the interface channel is supported solely by the small electrochemical potential difference between the two QH regions. As the interface potential is gentle, with negligible drift-velocity contribution, the plasmon velocity should be dominated by the Coulomb interaction. More importantly, random impurity potentials in both QH regions influence the interface mode, which must be considered when designing a high-quality channel.

\begin{figure}%[b]
\includegraphics[width=8.5 cm]{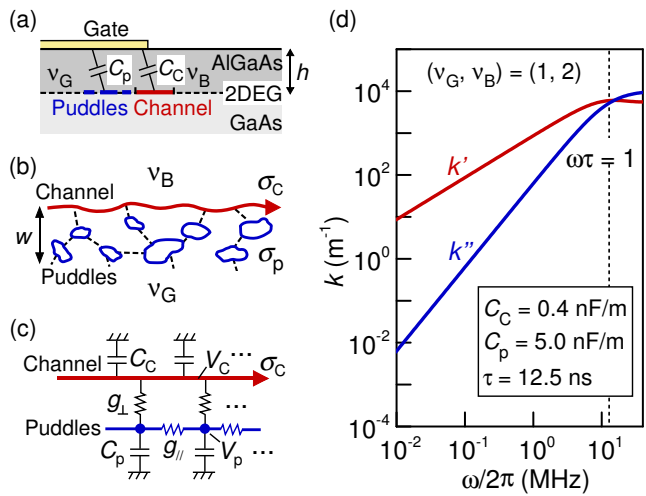}
\caption{Circuit model for the channel-puddle coupling. (a) Schematic cross-section around the interface channel formed between two QH states with filling factor $\nu_{\mathrm{B}}$ in the bulk of ungated region and $\nu_{\mathrm{G}}$ in the gated region. The interface channel and charge puddles formed under the gate coupled to the gate with a geometric capacitance $C_{\mathrm{C}}$ and $C_{\mathrm{p}}$, respectively. (b) Schematic illustration of the puddle array (blue closed curves) present in the gated region with a width $w$ from the channel and a local conductance $\sigma_{\mathrm{p}}$. (c) Distributed circuit model describing the coupling of the channel with the diffusive charge motion in the puddles. (d) Calculated wavenumber $k'$ and decay rate $k''$ as a function of frequency $\omega$ using representative parameters: $C_{\mathrm{C}}$ = 0.4 nF/m, $C_{\mathrm{p}}$ = 5 nF/m, and $\tau$ = 12.5 ns for ($\nu_{\mathrm{G}}$, $\nu_{\mathrm{B}}$) = (1, 2) in our device. Vertical dashed line marks the condition  $\omega\tau$ = 1.}
\end{figure}

In this study, we investigate the interface plasmon mode in both integer and fractional QH regimes. First, we introduce a distributed circuit model to describe how chiral plasmon transport can be influenced by diffusion processes in charge puddles. Then, a single interface channel is experimentally defined by preparing two QH states in the gated and ungated regions of an AlGaAs/GaAs heterostructure. Plasmon transport was investigated by exciting a charge wave packet in the channel and detecting it with a time-resolved charge detector. The obtained plasmon waveform is delayed and broadened during transport, from which the velocity and diffusion constant are evaluated with the model. A significant reduction in velocity accompanied by broadening of the wave packet is observed when the QH state under the gate is conductive even slightly. This can be overcome by placing a slightly conductive state in the ungated region. Indeed, a fractional interface channel with a reasonably fast velocity is obtained with a fractional state in the ungated region and an integer state in the gated region. Such high-quality interface channels are desirable for transporting fractional charges.

\section{Distributed circuit model for interface channel}
We consider an interface channel formed in an AlGaAs/GaAs heterostructure partially covered with a metal gate biased at an appropriate gate voltage, as shown in Fig. 1(a) \cite{Lin20,Khaetskii94}. Two QH states are formed with Landau-level filling factors, namely $\nu_{\mathrm{B}}$ in the bulk of the ungated region and $\nu_{\mathrm{G}}$ in the gated region. The interface channel has a conductance of $\sigma_{\mathrm{C}} = \Delta\nu e^2/h$ with $\Delta\nu= |\nu_{\mathrm{B}}-\nu_{\mathrm{G}}|$. In practice, the electrostatic potential of each QH region is spatially fluctuated by remote impurities or other factors, and thus charge puddles exist everywhere with excess or deficit of the filling factors \cite{Zhitenev00}. These conducting charge puddles provide a microscopic model for dissipative bulk conduction in the presence of disorders. Although the bulk of the QH states  at integer and fractional fillings can be insulating with vanishing longitudinal conductivity owing to the Anderson localization \cite{Huckestein95}, there may be conducting charge puddles locally coupled to the channel in each QH region. Although these puddles should not alter the standard dc conductance as far as the bulk is insulating, they can influence plasmon transport. We consider conductive puddles only in the QH region under the gate, as shown in Fig. 1(b), as these puddles have a large capacitance to the gate. A significant fraction of charge in the plasmon mode can be trapped by the puddle capacitors, which makes the plasmon mode dispersive and dissipative as shown in the following analysis. Similar conductive puddles may exist in the ungated QH region but should have only a minor effect on the plasmon mode owing to their small capacitance to the gate. We assume that the overall ensemble of the conductive puddles is characterized by an effective local longitudinal conductivity $\sigma_{\mathrm{p}} (\ll\sigma_{\mathrm{C}})$ for effective width $w$.

Such disorder effect can be understood with a distributed circuit model, as shown in Fig. 1(c), where only an array of puddles (solid circles) is considered for simplicity \cite{Lin20,Safi99,Hashisaka12,Hashisaka13,Kumada14,Fujisawa21}. The puddles are coupled to the channel with conductance $g_{\mathrm{\perp}}=\sigma_{\mathrm{p}}/w$ and to the neighboring puddles with conductance $g_{\mathrm{\parallel}}=\sigma_{\mathrm{p}}w$. The channel and puddles have capacitances $C_{\mathrm{C}}$ and $C_{\mathrm{p}}$, respectively, to the ground (or gates). These elements are defined as distributed elements with proper units for a unit length. While coupling capacitances may be considered in parallel to $g_{\mathrm{\perp}}$ and $g_{\parallel}$, these capacitances can be absorbed in $C_{\mathrm{C}}$, $C_{\mathrm{p}}$, and $g_{\mathrm{\perp}}$, and thus neglected, in the long-wavelength limit. The model describes how the chiral plasmon mode in a 1D channel is coupled to the diffusive motion in the puddle array. Using the current conservation law, we obtain a coupled wave equation:
\begin{align}
C_{\mathrm{C}} \frac{\partial V_{\mathrm{C}}}{\partial t} &= - \sigma_{\mathrm{C}} \frac{\partial V_{\mathrm{C}}}{\partial x}-g_{\mathrm{\perp}}(V_{\mathrm{C}}-V_{\mathrm{p}}) \\
C_{\mathrm{p}} \frac{\partial V_{\mathrm{p}}}{\partial t} &= g_{\mathrm{\parallel}} \frac{\partial^2 V_{\mathrm{p}}}{\partial x^2}+g_{\mathrm{\perp}}(V_{\mathrm{C}}-V_{\mathrm{p}})  \label{second}
\end{align}
for the voltages $V_{\mathrm{C}}(x, t)$ of the channel and $V_{\mathrm{p}}(x, t)$ of the puddles. Here, we considered chiral current ($\sigma_{\mathrm{C}}V_{\mathrm{C}}$) in the channel, nonchiral current ($g_{\mathrm{\parallel}}\partial V_{\mathrm{p}}/\partial x$) in the puddle array, and scattering current [$g_{\mathrm{\perp}}(V_{\mathrm{C}}-V_{\mathrm{p}})$] between the channel and the puddles. We find a solution in the form of exp$[i(kx - \omega t)]$ with frequency $\omega$ and complex $k = k' + ik''$, where the real part $k'$ and the imaginary part $k''$ describe the wave number and decay rate, respectively. We investigate the interface plasmon mode weakly coupled to the diffusive modes by neglecting the $k^2$ and $k^3$ terms in the secular equation. We focus on the long-wavelength limit with wavelength $\lambda \gg w$ and the weak-scattering limit with $g_{\mathrm{\perp}}g_{\mathrm{\parallel}}\ll\sigma_{\mathrm{C}}^2$ (i.e., $ \sigma_{\mathrm{p}}\ll\sigma_{\mathrm{C}}$). In these limits, $k'$ and $k''$ are approximately given by
\begin{align}
k' &\cong \frac{C_{\mathrm{C}}+C_{\mathrm{p}}(1+\omega^2\tau^2)^{-1} }{\sigma_{\mathrm{C}}}\omega \\
k'' &\cong \frac{\omega^2\tau C_{\mathrm{p}}}{\sigma_{\mathrm{C}}(1+\omega^2\tau^2)}
\end{align}
where $\tau = C_{\mathrm{p}}/g_{\mathrm{\perp}}$ is the effective charging time of the puddles. As shown in the dispersion relation in Fig. 1(d), the dissipation is significant at a higher frequency ($\omega \gtrsim 1/\tau$), and thus we expect the plasmon transport to be visible only in the range of $k'' \lesssim k'$ (i.e., $\omega\tau \lesssim 1$). As the actual charging time is distributed in the ensemble of the puddles, the use of this model with the effective $\tau$ should be restricted to the region of $\omega\tau \ll 1$. In the low-frequency regime at $\omega\tau \ll 1$, the charge wave propagates at a constant velocity $v_{\mathrm{C}} = \sigma_{\mathrm{C}}/(C_{\mathrm{C}} + C_{\mathrm{p}})$ with a decay length $l = \sigma_{\mathrm{C}}/\omega^2\tau C_{\mathrm{p}}$. When an initial wave packet is introduced to the channel, the packet is broadened during propagation. By neglecting the $\omega^2\tau^2$ term, the wave packet will be broadened in a Gaussian form
\begin{align}
V_{\mathrm{C}} \varpropto \mathrm{exp}[-(t-x/v_{\mathrm{C}})^2/4Dx]
\end{align}
with a diffusion constant $D = \tau C_{\mathrm{p}}/\sigma_{\mathrm{C}}$. We use $v_{\mathrm{C}}$ and $D$ to characterize the puddles in the system. When a finite $\omega^2\tau^2$ term is taken into consideration, the mode becomes dispersive and the wave packet shows asymmetric broadening with a long tail. The asymmetric broadening can be simulated using Eqs. (3) and (4) in the frequency domain or by numerically integrating Eqs. (1) and (2).

It should be noted that the above model can be applied to the case where the bulk is slightly conductive. Based on the theory of the edge magnetoplasmon mode with a semi-classical treatment \cite{Volkov88,Aleiner94,Johnson03}, the charge of mode is scattered into the bulk but still confined near the edge unless the charge reaches the opposite edge of the sample. This penetration length can be considered as $w$ in our model. Then, the velocity reduction and broadening can be understood with the model even for QH states with nonvanishing longitudinal conductance, like fractional states in the following experiment.

\section{Defining interface channels}
Figure 2(a) shows a schematic view of the device, which was fabricated from a standard GaAs/AlGaAs heterostructure with a 2DEG located $h$ = 100 nm below the surface \cite{Lin20}. A perpendicular magnetic field $B$ was applied to prepare a QH state with a filling factor $\nu_{\mathrm{B}}$ in the bulk. By applying an appropriate gate voltage $V_{\mathrm{g}}$ to the large metal gate (yellow region), another QH region with a different filling factor $\nu_{\mathrm{G}}$ was prepared in the gated region so that the interface channel was formed along the perimeter of the gate. The electron density was 1.85$\times$10$^{11}$ cm$^{-2}$ in the dark and 2.07$\times$10$^{11}$ cm$^{-2}$ after light irradiation at low temperature. All measurements were carried out at $\sim$ 50 mK and in a magnetic field up to 12 T.

\begin{figure}[t]
\includegraphics[width=8.3 cm]{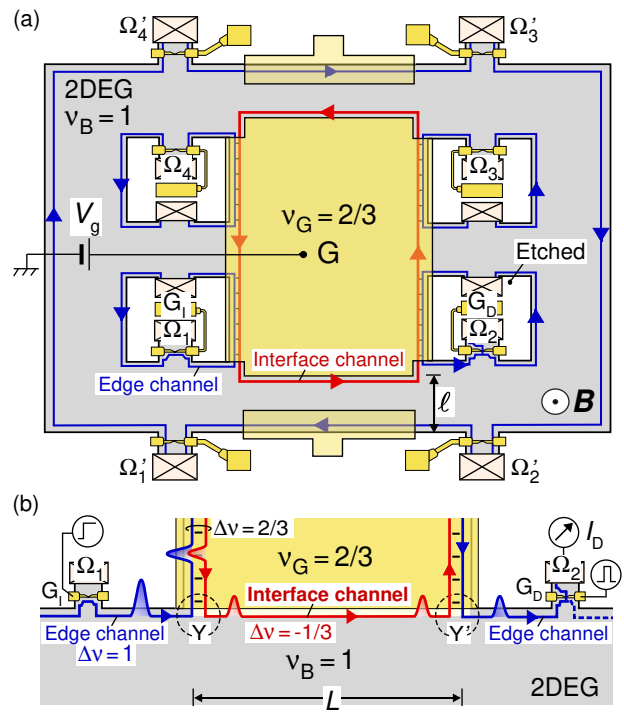}
\caption{Device and experimental setup for the time-resolved measurement. (a) Schematic view of the device. An interface channel is formed around the perimeter of the central gate (yellow region) by applying a voltage $V_{\mathrm{g}}$ with $\nu_{\mathrm{G}}$ (= 2/3) in the gated region set in the bulk $\nu_{\mathrm{B}}$ (= 1). The interface channel is separated from the outer edge by a distance of $\ell$ = 100 $\mu$m and can be connected by scattering with edge channels emanating from ohmic contacts in the Corbino geometry. (b) Setup for the time-resolved measurement. An initial charge packet is excited by applying a voltage step to the gate G$_{\mathrm{I}}$. The charge packet transmitted through the junction Y, interface channel ($L$ = 420 $\mu$m), and junction Y' can be detected by applying a voltage pulse to the gate G$_{\mathrm{D}}$.}
\end{figure}

Low-frequency transport through the interface channel was investigated using a Corbino-type device with four ohmic contacts [labeled $\Omega_1$, $\Omega_2$, $\Omega_3$, and $\Omega_4$ in Fig. 2(a)], which are attached to the inner edges of the hollowed 2DEG region \cite{Lin19}. A four-terminal measurement of the interface channel is made by probing the voltage difference $V_{\mathrm{xx}}$ between $\Omega_1$ and $\Omega_2$ under current $I_3$ flowing from $\Omega_4$ to $\Omega_3$, as shown in the inset of Fig. 3(a). Figure 3(a) shows the color-scale plot of $V_{\mathrm{xx}}$ as a function of the gate voltage $V_{\mathrm{g}}$ and magnetic field $B$. The overall patterns can be understood with a variation of $\nu_{\mathrm{B}}$ in the bulk shown by horizontal lines (black) and $\nu_{\mathrm{G}}$ under the gate shown by inclined lines (red). Vanishing $V_{\mathrm{xx}}$, which appears as white regions, is seen around the intersections of two lines with different $\nu_{\mathrm{G}}$ and $\nu_{\mathrm{B}}$ (for $\nu_{\mathrm{G}}$ = $\nu_{\mathrm{B}}$, there is no interface channel). For example, at ($\nu_{\mathrm{G}}$, $\nu_{\mathrm{B}}$) = (1, 2) marked by the square, the interface channel with $\Delta\nu$ = 1 (spin-down branch of the lowest Landau level) connects the ohmic contacts, while the other channels for the spin-up branch are isolated from each other, as shown in Fig. 3(c). Transport through this single interface channel was confirmed by observing two-terminal conductance $G_{\mathrm{2w}}$ between $\Omega_1$ and $\Omega_2$ with other ohmic contacts floating. As shown in Fig. 4(c), $G_{\mathrm{2w}}$ shows a clear plateau $G_{\mathrm{2w}}$ = $e^2/h$ around $V_{\mathrm{g}}$ = $-$0.13 V corresponding to ($\nu_{\mathrm{G}}$, $\nu_{\mathrm{B}}$) = (1, 2). Bulk scattering was confirmed to be negligible in both gated and ungated regions from the vanishing four-terminal resistance $R_{\mathrm{xx}} = V_{\mathrm{xx}}/I_3 \sim 0$, as shown in Fig.\,4(d).

\begin{figure}[t]%\begin{figure*}[t]
\includegraphics[width=8.5 cm]{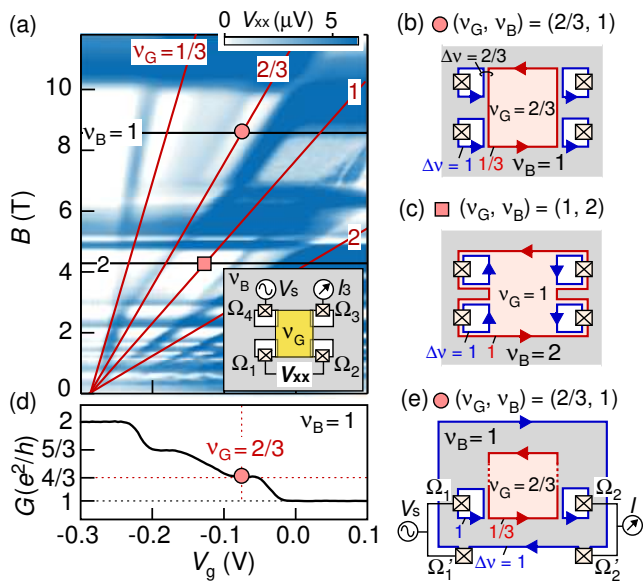}
\caption{Low-frequency characteristics of the interface channel. (a) Color plot of $V_{\mathrm{xx}}$ measured as a function of gate voltage $V_{\mathrm{g}}$ and magnetic field $B$. The four-terminal measurement setup is shown in the inset: A source voltage $V_{\mathrm{s}}$ = 30 $\mu$V at frequency 37 Hz is applied between $\Omega_4$ and $\Omega_3$, voltage $V_{\mathrm{xx}}$ is measured with $\Omega_1$ and $\Omega_2$, and current $I_3$ at $\Omega_3$ is monitored to obtain the resistance $R_{\mathrm{xx}}$ = $V_{\mathrm{xx}}$/$I_3$. The channel structures at conditions marked with solid symbols are sketched in (b, c). (b) Channel structure at $\nu_{\mathrm{G}}$ = 2/3 and $\nu_{\mathrm{B}}$ = 1, where the interface channel $\Delta\nu$ = 1/3 is connected to the integer channel $\Delta\nu$ = 1 through charge scattering. (c) Channel structure at $\nu_{\mathrm{G}}$ = 1 and $\nu_{\mathrm{B}}$ = 2, where the interface channel $\Delta\nu$ = 1 (spin-down branch of the lowest Landau level) connected the ohmic contacts and the channels of spin-up branch are isolated. (d) $V_{\mathrm{g}}$-dependence of two-terminal conductance $G$ measured at $\nu_{\mathrm{B}}$ = 1. (e) Measurement setup for obtaining $G$ and channel structure for the condition marked by the solid circle in (d).}
\end{figure}%end{figure*}

The connection between the interface channel and the ohmic contacts is not straightforward in some cases. For example, at ($\nu_{\mathrm{G}}$, $\nu_{\mathrm{B}}$) = (2/3, 1), the conductance of the interface channel is fractional ($\Delta\nu$ = 1/3) and that of the edge channel emanating from the ohmic contact is integral ($\Delta\nu$ = 1). They are bound to form a composite $\Delta\nu$ = 2/3 channel, where $\Delta\nu$ = 1 and 1/3 channels are counterpropagating in proximity \cite{MacDonald90}, as illustrated in Fig. 3(b). Therefore, transport through the interface channel requires tunneling and equilibration inside the composite channel \cite{Kane94}. Because the $\Delta\nu$ = 2/3 channel in our device ($\sim$ 300 $\mu$m) is longer than the typical equilibration length (about 10 $\mu$m) to reach the equal electrochemical potential of the two channels \cite{Lin19,Grivnin14}, the counter-propagating channels must be fully equilibrated. This was confirmed by observing that the two-terminal conductance $G_{\mathrm{2w}}$ exhibits a plateau at $e^2/3h$, as shown in Fig. 5(d). The $R_{\mathrm{xx}}$ data in Fig. 5(e) show a minimum at around ($\nu_{\mathrm{G}}$, $\nu_{\mathrm{B}}$) = (2/3, 1). However, the small but finite $R_{\mathrm{xx}}$ that remains even at the minimum influences the plasmon transport, as shown in plasmon waveforms.

In the case of ($\nu_{\mathrm{G}}$, $\nu_{\mathrm{B}}$) = (2/3, 1), it is worth to noting that the interface channel ($\Delta\nu$ = 1/3) is well isolated from the $\Delta\nu$ = 1 channel along the outer edge of the device. This can be regarded as an artificial realization of the hierarchical edge structure without equilibration, for which two-terminal conductance of $4e^2/3h$ was anticipated \cite{MacDonald90,Kane94}. To test this idea, an effective two-terminal conductance measurement was performed by connecting the inner and outer ohmic contacts ($\Omega_1$ - $\Omega_1'$, and $\Omega_2$ - $\Omega_2'$) with all other ohmic contacts floating, as shown in Fig. 3(e). The conductance shows a clear plateau at $4e^2/3h$ as a function of $V_{\mathrm{g}}$, when the fractional QH state ($\nu_{\mathrm{G}}$ = 2/3) is formed at $V_{\mathrm{g}}$ = $-$0.08 V, as marked by the red circle in Fig. 3(d). This also ensures the formation of a single isolated interface channel ($\Delta\nu$ = 1/3) in our device.

\section{Plasmon measurement}
\subsection{Time-resolved measurement scheme}
To evaluate the plasmon transport, we employed a time-resolved waveform measurement scheme. As shown in Fig. 2(b) for ($\nu_{\mathrm{G}}$, $\nu_{\mathrm{B}}$) = (2/3, 1), the injection gate G$_{\mathrm{I}}$ is prepared with being fully depleted underneath by applying a sufficiently large negative static voltage ($-$300 mV). The addition of a voltage step ($V_{\mathrm{I}}$ = 15 mV) to the injection gate G$_{\mathrm{I}}$ further depletes nearby regions, and the depleted electrons travel as a pulsed charge packet in the edge channel ($\Delta\nu$ = 1). This charge packet propagates as an edge mode along the perimeter of the hollow before entering the interface channel. To avoid possible nonlinear effects \cite{Zhitenev94}, the induced rf current is kept at a low level ($\sim 0.1$ - 10 nA) which is comparable to or smaller than the current ($\sim 1$ nA) in the low-frequency $R_{\mathrm{xx}}$ measurements.

The connection between the edge channel and the interface channel can be understood by considering the transmission at the junction (Y junction) of the three channels ($\Delta\nu$ = $\nu_{\mathrm{B}}$, $\nu_{\mathrm{G}}$, and $|\nu_{\mathrm{B}} - \nu_{\mathrm{G}}|$) \cite{Lin20}. One third of the incident charge on the $\Delta\nu$ = 1 channel goes to the $\Delta\nu$ = 1/3 channel, while the remainder goes to the composite $\Delta\nu$ = 2/3 channel and is absorbed in a grounded ohmic contact. After traveling through the interface channel ($\Delta\nu$ = 1/3) of length $L$ = 420 $\mu$m, the charge packet is transferred to another edge channel through the other junction (Y').

The resultant charge packet is detected by applying a voltage pulse ($V_{\mathrm{D}}$ = 20 mV) of width $t_{\mathrm{w}}$ = 0.08 - 260 ns to the detection gate G$_{\mathrm{D}}$ with a controlled time delay $t_{\mathrm{d}}$ from the injection voltage step. The waveform of the charge packet can be obtained from the $t_{\mathrm{d}}$ dependence of dc current $I_{\mathrm{D}}$ at $\Omega_2$. Experimental details of this scheme can be found elsewhere \cite{Hashisaka17,Kamata14,Lin20}. Because the edge mode without metal on top has much faster charge velocity as compared to the interface mode with a nearby metal \cite{Kumada11}, we ignore the time-of-flight in this short ungated edge channel. The time origin of $t_{\mathrm{d}}$ was determined from the peak position of a reference trace taken at zero magnetic field, where the wave packet propagates in 2DEG at much faster velocity ($\sim 10^7$ m/s) \cite{Kumada11}. The delay of the peak position at high field can be used to determine the charge velocity, which has been well characterized for the general edge plasmon modes \cite{Ashoori92,Ernst96,Kamata10}.

As shown in Fig. 4(a), a clear peak with a narrow width of $\sim 0.45$ ns, comparable to the earlier reports \cite{Kumada11,Kamata09}, shows a reasonably high time resolution obtained with $t_{\mathrm{w}}$ = 0.08 ns. In our previous report, we showed that the charge in the wave packet was fractionalized at the Y junctions with a quantized ratio \cite{Lin20}. We focus on the plasmon transport in the interface channel in this paper. The scheme can be applied to interface channels of other realizations ($\nu_{\mathrm{B}} \neq \nu_{\mathrm{G}} > 0$) as well as to edge channels with full depletion under the gate ($\nu_{\mathrm{B}} \neq \nu_{\mathrm{G}} = 0$). This allows direct comparison between the interface and edge modes.

\begin{figure}%[t]
\includegraphics[width=8.5 cm]{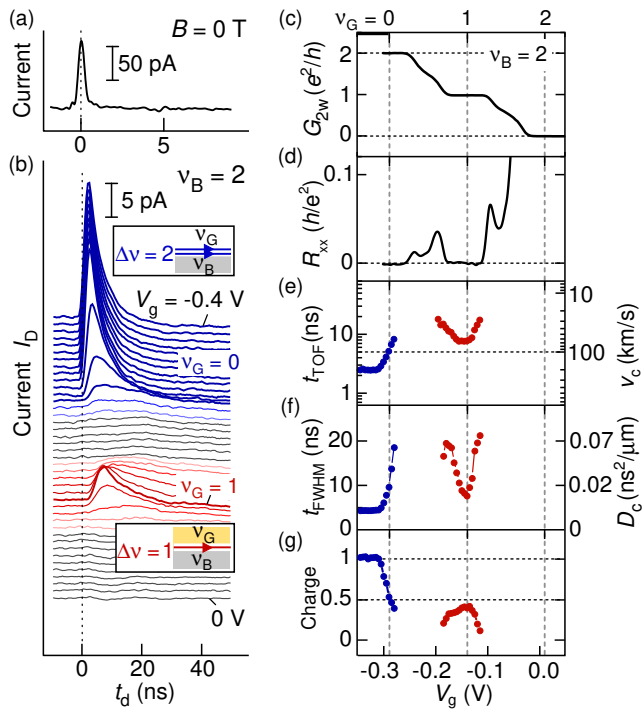}
\caption{Interface plasmon mode in integer QH regime. (a) Reference waveform measured at $B$ = 0 T. The time origin is determined by the peak position. (b) Waterfall plot of current $I_{\mathrm{D}}$ as a function of delay time $t_{\mathrm{d}}$ for various $V_{\mathrm{g}}$ values from 0 V (the bottom trace at $\nu_{\mathrm{G}} \cong 2$) to $-$0.4 V (the top trace at $\nu_{\mathrm{G}}$ = 0) with step 0.01 V obtained at $B$ = 4.2 T ($\nu_{\mathrm{B}}$ = 2). These waveforms are detected with a pulse width $t_{\mathrm{w}}$ = 0.15 ns and repetition time $t_{\mathrm{rep}}$ = 0.15 $\mu$s. The edge channel at $\nu_{\mathrm{G}}$ = 0 and interface channel at $\nu_{\mathrm{G}}$ = 1 are illustrated in the respective insets. (c-g) $V_{\mathrm{g}}$ dependence of conductance $G_{\mathrm{2w}}$ in (c), $R_{\mathrm{xx}}$ in (d), $t_{\mathrm{TOF}}$ in (e), $t_{\mathrm{FWHM}}$ in (f), and charge normalized by the value ($\sim$ 530$e$) at $V_{\mathrm{g}}$ = $-$0.35 V in (g). The charge velocity $v_{\mathrm{c}}$ and diffusion constant $D_{\mathrm{c}}$ are shown on the right axes of (e) and (f), respectively.}
\end{figure}

\subsection{Interface mode in the integer QH regime}
We first investigate the plasmon modes in the integer QH regime. Figure 4(b) shows the charge waveform in the current $I_{\mathrm{D}}$ as a function of delay time $t_{\mathrm{d}}$, obtained for various gate voltage $V_{\mathrm{g}}$ at $B$ = 4.2 T with $\nu_{\mathrm{B}}$ = 2. The large signal at $\nu_{\mathrm{G}}$ = 0 (full depletion under the gate at $V_{\mathrm{g}} < -$0.28 V, shown by the blue traces) is attributed to the propagation in the edge plasmon mode along the $\Delta\nu = |\nu_{\mathrm{B}} - \nu_{\mathrm{G}}|$ = 2 edge channel, as illustrated in the upper inset. The small but clear signal at around $\nu_{\mathrm{G}}$ = 1 ($V_{\mathrm{g}} \sim -$0.13 V, shown by the red traces) indicates the propagation through the interface plasmon mode in the $\Delta\nu$ = 1 channel, as shown in the lower inset. The wave packet passing through the interface channel ($\nu_{\mathrm{G}}$ = 1) is delayed and slightly broadened as compared to that through the edge channel ($\nu_{\mathrm{G}}$ = 0). At other gate voltages including $V_{\mathrm{g}} \sim$ 0 V corresponding to $\nu_{\mathrm{G}}$ = 2 (the lowest trace), very weak or almost no signal is detected as no well-defined channels are formed between the injector and the detector.

We extracted the representative time of flight $t_{\mathrm{TOF}}$ from the peak position and the full width at half maximum $t_{\mathrm{FWHM}}$ of the peak, as shown in Figs. 4(e) and 4(f), respectively. These characteristics are compared with the two-terminal conductance $G_{\mathrm{2w}}$ shown in Fig. 4(c) and the longitudinal resistance $R_{\mathrm{xx}}$ in Fig. 4(d). For the interface mode ($\Delta\nu$ = 1 at $\nu_{\mathrm{G}}$ = 1), shown by red symbols in Figs. 4(e) and 4(f), both $t_{\mathrm{TOF}}$ and $t_{\mathrm{FWHM}}$ become minimum at the center of the $G_{\mathrm{2w}}$ = $e^2/h$ plateau and the vanishing $R_{\mathrm{xx}}$ ($V_{\mathrm{g}} \sim -$0.13 V). Interestingly, both $t_{\mathrm{TOF}}$ and $t_{\mathrm{FWHM}}$ increase rapidly when $\nu_{\mathrm{G}}$ deviates only slightly from 1, despite that no measurable changes are seen in $G_{\mathrm{2w}}$ and $R_{\mathrm{xx}}$. This is in contrast to other studies of the edge plasmon mode, where the velocity and width can properly scale with the conductivity \cite{Grodnensky91,Ashoori92,Talyanskii94}. Based on the model described in Sec. II, the charge waveform probes the local puddles that are located under the gate and effectively coupled to the channel. The charge velocity $v_{\mathrm{c}} = L/t_{\mathrm{TOF}}$ and the diffusion constant $D_{\mathrm{c}} = t_{\mathrm{FWHM}}^2/16(\ln2)L$ are shown on the right axes of Figs. 4(e) and 4(f) as a guide, while the values might be influenced by the asymmetric broadening.

A similar analysis can be made for the edge channel formed at $V_{\mathrm{g}}$ = $-$0.30 $\sim$ $-$0.28 V, where $t_{\mathrm{TOF}}$ and $t_{\mathrm{FWHM}}$ increase rapidly when $\nu_{\mathrm{G}}$ increases only slightly above 0, where puddles appear under the gate. The data for the edge mode ($\Delta\nu$ = 2) can be quantitatively compared with that for the interface mode ($\Delta\nu$ = 1) by noting that both $t_{\mathrm{TOF}}$ and $t_{\mathrm{FWHM}}$ are inversely proportional to the channel conductance $\sigma_{\mathrm{C}}$ ($\propto \Delta\nu$). Namely, the $t_{\mathrm{TOF}}$ and $t_{\mathrm{FWHM}}$ values increasing with $\nu_{\mathrm{G}}$ at $\nu_{\mathrm{G}} \gtrsim$ 1 are close to double those increasing at $\nu_{\mathrm{G}} \gtrsim$ 0. In the case of the edge mode, the puddles under the gate can be completely eliminated by applying sufficiently negative $V_{\mathrm{g}}$ ($< -$0.3 V), where $t_{\mathrm{TOF}}$ and $t_{\mathrm{FWHM}}$ decrease further due to the reduction of $C_{\mathrm{C}}$ and $C_{\mathrm{p}}$ \cite{Kamata10}. However, for the interface mode, the puddles are always present even at integer filling $\nu_{\mathrm{G}}$ = 2, and electron (hole) puddles develop for $\nu_{\mathrm{G}} >$ 2 ($\nu_{\mathrm{G}} <$ 2).

While we observed significant broadening for the interface mode, the charge must be confined within the edge as far as $R_{\mathrm{xx}}$ remains zero. The charge in the wave packet, which is evaluated by the area of the current peak in Fig. 4(b), is plotted as a function of $V_{\mathrm{g}}$ in Fig. 4(g), where the vertical axis is normalized by the value at $V_{\mathrm{g}}$ = $-$0.35 V ($\nu_{\mathrm{G}}$ = 0). As compared to the charge for the edge mode ($\Delta\nu$ = 2) at $\nu_{\mathrm{G}}$ = 0, the charge for the interface mode ($\Delta\nu$ = 1) at $\nu_{\mathrm{G}}$ = 1 is approximately halved because only one channel is connected between the $\Delta\nu$ = 2 edges along the hollows. This ratio is not necessarily to be exactly 1/2 and should be determined by the charge distribution between the two channels along the hollow \cite{Hashisaka17}. Nevertheless, the ratio close to 1/2 indicates that the charge remains in the interface mode in the vicinity of $\nu_{\mathrm{G}}$ = 1 where $R_{\mathrm{xx}} \sim 0$. This is the signature of broadening associated with the local conductive region (the charge puddles). The charge in the charge wave packet decays rapidly when $R_{\mathrm{xx}}$ becomes finite, but this strong dissipative regime studied in previous works \cite{Grodnensky91,Ashoori92,Talyanskii94} is not within the scope of this paper.

\subsection{Interface mode in the fractional QH regime}
We can apply the model developed in Sec. II to the weak dissipative regime where the bulk $R_{\mathrm{xx}}$ is finite but small so that the charge is well confined near the edge. The model suggests that the puddles significantly influence the interface mode if the QH state under the gate is not completely insulating, as in the $\nu_{\mathrm{G}}$ = 2/3 case in our device. Figures 5(a) and 5(b) show the plasmon waveforms $I_{\mathrm{D}}(t_{\mathrm{d}})$ for various $V_{\mathrm{g}}$ ($0 \leq \nu_{\mathrm{G}} \leq 1$) at $B$ = 8.7 T ($\nu_{\mathrm{B}}$ = 1). As shown in Fig. 5(a), a sharp peak is resolved when an integer channel ($\nu_{\mathrm{G}}$ = 1) is formed with full depletion under the gate ($\nu_{\mathrm{G}}$ = 0 at $V_{\mathrm{g}} < -$0.32 V). No wave packet signal for fractional channels was detected with this short $t_{\mathrm{w}}$ (= 0.08 ns) and $t_{\mathrm{rep}}$ (= 32 ns), because the wave packet was broadened too much. Even in this situation, clear plasmon transport should appear at lower frequencies [$\omega < 1/\tau$ in Eqs. (3) and (4)]. This was confirmed in our experiment just by increasing the excitation and detection time to $t_{\mathrm{w}}$ = 260 ns and $t_{\mathrm{rep}}$ = 13 $\mu$s (the measurement frequency ranges from $\sim 1/t_{\mathrm{rep}}$ to $\sim 1/t_{\mathrm{w}}$), as shown in Fig. 5(b). With this board ``boxcar" window of $t_{\mathrm{w}}$, the sharp peak for $\Delta\nu$ = 1 is broadened into a rectangular shape as seen in the topmost trace at $V_{\mathrm{g}}$ = $-$0.3 V. When the fractional interface channel ($\Delta\nu$ = 1/3) is activated with $\nu_{\mathrm{G}}$ = 2/3 at $V_{\mathrm{g}}$ = $-$0.092 V, a clear charge wave packet is observed as shown by the red traces in Fig. 5(b). Similarly, the wave packet through another fractional interface channel ($\Delta\nu$ = 2/3) activated with $\nu_{\mathrm{G}}$ = 1/3 at $V_{\mathrm{g}}$ = $-$0.204 V is also resolved (the green traces). As this measured waveform is slightly influenced by the wide boxcar window, the actual waveform can be estimated by taking the deconvolution as shown in Fig. 5(c) for the representative cases at $\nu_{\mathrm{G}}$ = 0, 1/3, and 2/3. Significant time delay and broadening are clearly seen for the interface modes.

\begin{figure}[t]
\includegraphics[width=8.5 cm]{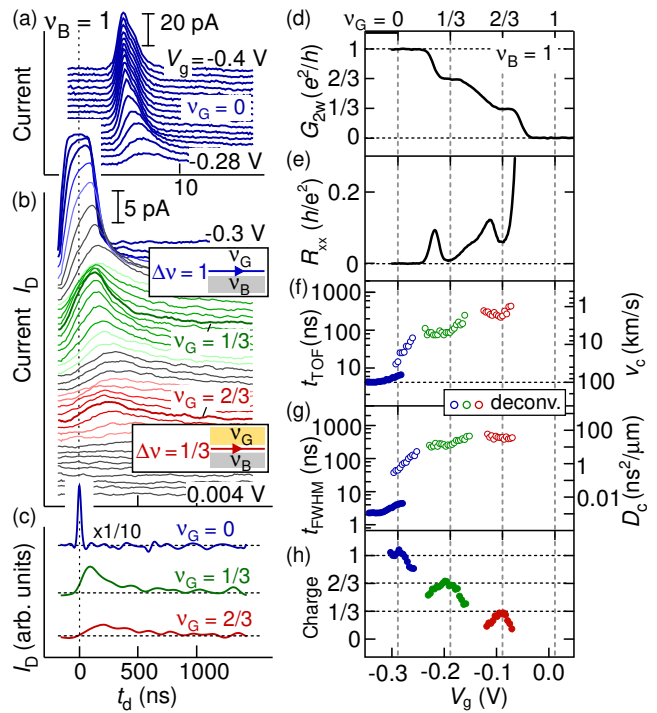}
\caption{Interface plasmon mode in fractional QH regime. (a)-(c) Waterfall plot of current $I_{\mathrm{D}}$ as a function of delay time $t_{\mathrm{d}}$ for various $V_{\mathrm{g}}$ obtained at $B$ = 8.7 T ($\nu_{\mathrm{B}}$ = 1). The data in (a) with $V_{\mathrm{g}}$ from $-$0.28 V to $-$0.4 V are measured with a short detector pulse width $t_{\mathrm{w}}$ = 0.08 ns and repetition time $t_{\mathrm{rep}}$ = 32 ns. The data in (b) with $V_{\mathrm{g}}$ from 0.004 V to $-$0.3 V are measured with $t_{\mathrm{w}}$ = 260 ns and $t_{\mathrm{rep}}$ = 13 $\mu$s. The edge channel at $\nu_{\mathrm{G}}$ = 0 and interface channel at $\nu_{\mathrm{G}}$ = 2/3 are illustrated in the respective insets. The deconvoluted waveform at $\nu_{\mathrm{G}}$ =0, 1/3, and 2/3 is shown in (c), obtained by deconvolution of the waveform in (b) with the 260 ns ``boxcar'' window. (d-h) $V_{\mathrm{g}}$ dependence of conductance $G_{\mathrm{2w}}$ in (d), $R_{\mathrm{xx}}$ in (e), $t_{\mathrm{TOF}}$ in (f), $t_{\mathrm{FWHM}}$ in (g), and charge normalized by the value ($\sim$ 8000$e$) at $V_{\mathrm{g}}$ = $-$0.3 V in (h). The charge velocity $v_{\mathrm{c}}$ and diffusion constant $D_{\mathrm{c}}$ are shown on the right axes of (f) and (g), respectively.}
\end{figure}

In the same way as in the integer case, $t_{\mathrm{TOF}}$, $t_{\mathrm{FWHM}}$, and the charge of the charge packets are plotted in Figs. 5(f), 5(g), and 5(h), respectively. Here, the values estimated from the deconvoluted waveforms are shown by open symbols. First the charge normalized by the value at $\nu_{\mathrm{G}}$ = 0 is found to be 2/3 and 1/3 when the fractional channels $\Delta\nu$ = 2/3 and 1/3 are formed, respectively. This means that the charge is conserved within the interface channel even though $R_{\mathrm{xx}}$ is finite \cite{Lin20}. This validates that the obtained $t_{\mathrm{TOF}}$ and $t_{\mathrm{FWHM}}$ can be analyzed with our model shown in Sec. II. The fractional interface channels have much slower velocities, $\sim$ 1.8\,km/s for the interface $\Delta\nu$ = 1/3 channel and  $\sim$ 4.4 km/s for the $\Delta\nu$ = 2/3 channel, as compared to $\sim$ 100 km/s for the $\Delta\nu$ = 1 edge channel.

The significantly different velocities can be related to the local conductivity $\sigma_{\mathrm{p}}$ due to the ensemble of puddles introduced in Sec. II. To see this, we made a crude estimate of $\sigma_{\mathrm{p}} = C_{\mathrm{p}}^2/D\sigma_{\mathrm{C}}$ by assuming $C_{\mathrm{p}} \gg C_{\mathrm{C}}$, which is justified when the velocity of the interface plasmons, $\sim \sigma_{\mathrm{C}}/(C_{\mathrm{p}} + C_{\mathrm{C}})$, is significantly lower than that of the edge plasmons, $\sim \sigma_{\mathrm{C}}/C_{\mathrm{C}}$, with no puddles under the gate. This $\sigma_{\mathrm{p}}$ can be compared with dc conductivity by ignoring the possible frequency dependence. For this comparison, we use the four-terminal conductance $g_{\mathrm{xx}}$ instead of the conductivity by ignoring the unknown geometrical factor (on the order of 1), where $g_{\mathrm{xx}}$ can be obtained as $g_{\mathrm{xx}}=R_{\mathrm{xx}}/(R_{\mathrm{xx}}^2+R_{\mathrm{xy}}^2 )\cong R_{\mathrm{xx}}(\nu_{\mathrm{G}}e^2/h)^2$ for sufficiently small $R_{\mathrm{xx}}$ ($\ll R_{\mathrm{xy}} = h/\nu_{\mathrm{G}}e^2$). We find comparable values for $\sigma_{\mathrm{p}}$ and $g_{\mathrm{xx}}$: $\sigma_{\mathrm{p}} \simeq 0.021e^2/h$ and $g_{\mathrm{xx}} \simeq 0.02e^2/h$ for the $\Delta\nu$ = 1/3 channel ($\nu_{\mathrm{G}}$ = 2/3), $\sigma_{\mathrm{p}} \simeq 0.007e^2/h$ and $g_{\mathrm{xx}} \simeq 0.002e^2/h$ for the $\Delta\nu$ = 2/3 channel ($\nu_{\mathrm{G}}$ = 1/3), and $\sigma_{\mathrm{p}} \simeq 0.005e^2/h$ and $g_{\mathrm{xx}} \ll 0.001e^2/h$ (noise level) for the $\Delta\nu$ = 1 channel ($\nu_{\mathrm{G}}$ = 1). This supports the validity of the model and suggests that smaller $\sigma_{\mathrm{p}}$ is preferred for small broadening of the wave packet.

The above analysis should be performed in the low-frequency limit ($\omega \tau \ll 1$) of our model (Sec. II). This is related to the choice of the boxcar window ($t_{\mathrm{w}}$). The effective charging time $\tau = C_{\mathrm{p}}w/\sigma_{\mathrm{p}}$ is estimated from the above $C_{\mathrm{p}}$ and $\sigma_{\mathrm{p}}$. Here, $w \simeq C_{\mathrm{p}} h/\varepsilon_{\mathrm{GaAs}}$ can be estimated from a parallel-plate capacitance approximation between the puddle and the gate. By considering the measurable frequency range of $\omega = 2\pi/t_{\mathrm{rep}} \sim \pi/t_{\mathrm{w}}$, $\omega\tau$ ranges 0.04 $\sim$ 1 for the $\Delta\nu$ = 1/3 channel ($\nu_{\mathrm{G}}$ = 2/3) and 0.03 $\sim$ 0.7 for the $\Delta\nu$ = 2/3 channel ($\nu_{\mathrm{G}}$ = 1/3). This indicates that the wave packet becomes visible by restricting ourselves in the low-frequency regime ($\omega \tau \lesssim 1$) with large $t_{\mathrm{w}}$.

The influence of charge puddles can be reduced by increasing the energy gap of the QH state in the gated region. This can be done by increasing $B$ and simultaneously increasing the electron density to maintain the same $\nu_{\mathrm{G}}$. Figure 6(a) shows several waveforms $I_{\mathrm{D}}(t_{\mathrm{d}})$ (solid circles) as well as their deconvolution (open circles) at fixed $\nu_{\mathrm{G}}$ = 2/3 with different $B$ and $V_{\mathrm{g}}$, while the state in the ungated region remains insulating in the range of 0.98 $< \nu_{\mathrm{B}} <$ 1.16. Under these conditions denoted by ($\nu_{\mathrm{G}}$, $\nu_{\mathrm{B}}$) = (2/3, $\sim$1), the wave packet becomes sharper and less delayed with increasing $B$.

The slow velocity can also be interpreted with weak Coulomb interaction screened by the gate for distributed charges in the puddles. Ultimately, the velocity might be decreased to the single-particle drift velocity $v_{\mathrm{d}} = E_{\mathrm{\perp}}/B$ determined by the perpendicular electric field $E_{\mathrm{\perp}}$ and magnetic field $B$. While we do not know typical $v_{\mathrm{d}}$ values for the interface channels, the data in Fig. 6(a) do not follow the $1/B$ dependence. This implies that the slow velocity is still dominated by the Coulomb interaction and can be understood with the geometric capacitances in the model.

\begin{figure}[t]
\includegraphics[width=8.5 cm]{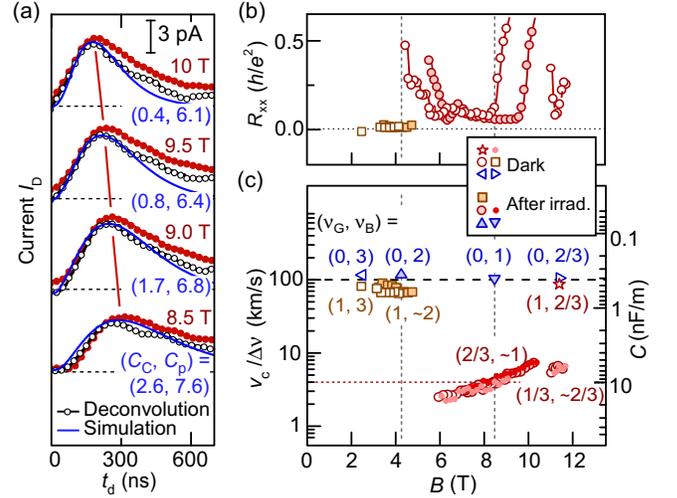}
\caption{Velocity of interface plasmon mode. (a) Waveforms (solid circles) and their deconvolution (open circles) at fixed $\nu_{\mathrm{G}}$ = 2/3 with different $B$ and $V_{\mathrm{g}}$, where $\nu_{\mathrm{B}}$ changes slightly around 1. Simulated waveforms using ($C_{\mathrm{C}}$, $C_{\mathrm{p}}$) values are plotted in solid lines, where $C$ (= $C_{\mathrm{C}}$ + $C_{\mathrm{p}}$) and $g_{\mathrm{\perp}}$ were obtained from $t_{\mathrm{TOF}}$ and $t_{\mathrm{FWHM}}$, respectively, and $C_{\mathrm{C}}$ is obtained from the fitting. (b),(c) $B$ dependence of resistance $R_{\mathrm{xx}}$ in (b), and normalized charge velocity $v_{\mathrm{c}}/\Delta\nu$ for several ($\nu_{\mathrm{G}}$, $\nu_{\mathrm{B}}$) conditions in (c). Data points are marked with circles for fractional interface channel at ($\nu_{\mathrm{G}}$, $\nu_{\mathrm{B}}$) = (2/3, $\sim$1), (1/3, $\sim$2/3) and star at (1, 2/3), quadrangles for integer interface channels at ($\nu_{\mathrm{G}}$, $\nu_{\mathrm{B}}$) = (1, 3) and (1, $\sim$2), triangles for edge channels at ($\nu_{\mathrm{G}}$, $\nu_{\mathrm{B}}$) = (0, 3), (0, 2), (0, 1), and (0, 2/3), and are filled with white for the device in the dark and light colors for the device after light irradiation. The velocity obtained from the deconvoluted waveforms are shown with small solid circles. The channel capacitance $C$ is shown on the right axis of (c). }
\end{figure}

\subsection{Normalized velocity}
We repeated such measurements under various conditions ($\nu_{\mathrm{G}}$, $\sim$$\nu_{\mathrm{B}}$). The $R_{\mathrm{xx}}$ from the four-terminal measurement and $v_{\mathrm{c}}$ evaluated from the plasmon measurement are summarized in Figs. 6(b) and 6(c), respectively. Here, the normalized velocity  $v_{\mathrm{c}}/\Delta\nu$ is plotted in Fig. 6(c), because the velocity increases in proportion to $\Delta\nu$ of the channel. The vertical axis is translated to the capacitance $C = \Delta\nu e^2/hv_{\mathrm{c}}$, as shown on the right axis. This capacitance can be understood as the total capacitance $C = C_{\mathrm{C}} + C_{\mathrm{p}}$ based on the puddle model in Sec. II. The normalized velocity of the edge channel with $\nu_{\mathrm{G}}$ = 0 is constant as shown by the blue triangles for ($\nu_{\mathrm{G}}$, $\nu_{\mathrm{B}}$) = (0, 3), (0, 2), (0, 1), and (0, 2/3), where no puddles are present under the gate. In contrast, the interface channels show smaller normalized velocities, as shown by the squares for ($\nu_{\mathrm{G}}$, $\nu_{\mathrm{B}}$) = (1, 3) and (1, $\sim$2) in the integer regime and circles for ($\nu_{\mathrm{G}}$, $\nu_{\mathrm{B}}$) = (2/3, $\sim$1) and (1/3, $\sim$2/3) in the fractional regime. Note that the normalized velocity for ($\nu_{\mathrm{G}}$, $\nu_{\mathrm{B}}$) = (2/3, $\sim$1) increases with $B$. This is consistent with the gradual reduction of $R_{\mathrm{xx}}$ with increasing $B$, as shown by the circles in Fig. 6(b). The steep increase in $R_{\mathrm{xx}}$ seen at both ends of the ($\nu_{\mathrm{G}}$, $\nu_{\mathrm{B}}$) = (2/3, $\sim$1) data is attributed to the backscattering in the ungated region, as $\nu_{\mathrm{B}}$ is deviated greatly from 1. However, no visible change in the velocity is seen even when the scattering in the ungated region sets in. This supports the validity of our model in which the gate capacitance of the puddles plays an important role in determining the velocity and broadening of the plasmon.

Among the various conditions for our devices, the fractional interface channel $\Delta\nu$ = 1/3 at ($\nu_{\mathrm{G}}$, $\nu_{\mathrm{B}}$) = (1, 2/3) shows reasonably fast plasmon velocity, as shown by the star in Fig. 6(c). This was measured with positive $V_{\mathrm{g}}$ = 0.215 V to prepare a $\nu_{\mathrm{G}}$ = 1 QH state under the gate of a similar device, as described in Ref. 20. This contrasts with the slow velocities obtained when $\nu_{\mathrm{G}}$ and $\nu_{\mathrm{B}}$ are swapped, i.e., for ($\nu_{\mathrm{G}}$, $\nu_{\mathrm{B}}$) = (2/3, $\sim$1). In other words, placing the highly insulating $\nu$ = 1 region under the gate and the poorly insulating $\nu$ = 2/3 region away from the gate significantly reduces the capacitance of the puddles. As the data for ($\nu_{\mathrm{G}}$, $\nu_{\mathrm{B}}$) = (1, 2/3) were taken at higher $B$, the larger gap of the 2/3 state (and hence smaller $R_{\mathrm{xx}}$) could be partially responsible for the higher velocity. This should have a minor effect, as in Fig. 6(c), because extrapolating the $v_{\mathrm{c}}/\Delta\nu$ data for ($\nu_{\mathrm{G}}$, $\nu_{\mathrm{B}}$) = (2/3, $\sim$1) to higher $B$ does not reach the value for (1, 2/3). As the normalized velocity at ($\nu_{\mathrm{G}}$, $\nu_{\mathrm{B}}$) = (1, 2/3) is close to the value obtained with the integer interface channels as well as general edge channels, the channel is barely affected by the puddles in the gated region. This suggests the interface fractional channel at (1, 2/3) has a similar small capacitance $C \approx C_{\mathrm{C}}$, in stark contrast to the large $C \approx C_{\mathrm{p}}$ ($\gg C_{\mathrm{C}})$ at (2/3, $\sim$1) that is dominated by the puddle capacitance. Such clean fractional channels are highly desirable for transporting fractional charges.

\subsection{Asymmetric waveform}
Most of the charge waveforms presented here are broadened asymmetrically with a longer tail. This can be understood as retardation due to the puddles. By using an initial wave packet in a Gaussian form of width $\tau$, we calculated the final waveform after propagation by numerically integrating Eqs. (1) and (2). This reproduces the asymmetric broadening, as shown by the solid lines in Fig. 6(a), where $C_{\mathrm{C}}$ is adjusted to fit the curve to the data. Our simple model considers puddles with representative conductance $\sigma_{\mathrm{p}}$ and capacitance $C_{\mathrm{p}}$. In reality, more puddles with different parameters are present around the channel. Some puddles with higher conductance to the channel can contribute to increasing $C_{\mathrm{C}}$ in the model. This could be the reason why $C_{\mathrm{C}}$ increases with decreasing $B$ in the fitting. Other puddles with lower conductance (longer retardation) were neglected in our model, which could be the reason for the remaining deviation in the long tail. Puddles in the ungated region with small capacitance can be included in the model for a better understanding. While the model can be improved by considering the variation of the puddles, the presented model captures most of the experimental features including the asymmetric broadening.

It should be emphasized that our model can also be used to understand the dissipation of the conventional edge magnetoplasmon mode. The effects of disorder on the edge magnetoplasmon have been theoretically studied without considering the effect of the gate metal \cite{Johnson03}, which helped to understand the velocity reduction and waveform broadening observed at noninteger filling \cite{Kumada14,Tu18}. Our model reveals the crucial role of the gate-puddle capacitive coupling on the dissipation of the plasmon mode, which explains why these effects become more pronounced in gated samples \cite{Kumada11}.

\hspace*{\fill} \\

%\vspace{0cm}
%\setlength{\gnat}{1cm}

\section{Conclusions}
In summary, we have investigated the interface plasmon mode in both integer and fractional QH regimes using a time-resolved waveform measurement scheme. The obtained plasmon waveform is delayed and broadened, which can be well understood with a distributed circuit model describing the coupling of the plasmon mode with the diffusion process in the charge puddles. The mode is influenced more strongly by the puddles in the gated region than by those in the ungated region. Indeed, when a fractional state with a small energy gap is on the gated side, the fractional channel is subject to significant velocity reduction and broadening. Meanwhile, a clean fractional channel with a reasonably fast velocity is realized when placing the fractional state on the ungated side. Such high-quality interface fractional channels can be further utilized to transport fractional charges for studying the non-trivial fractional statistics \cite{Halperin84,Nakamura19,Bartolomei20,Nakamura20} and can be used to explore the many-body quantum dynamics in the fractional quantum edge state.

\vspace{0.5cm}

\begin{acknowledgements} The authors thank T. Hata and Y. Tokura for their beneficial discussions. This study was supported by Grants-in-Aid for Scientific Research No. JP15H05854, and No. JP19H05603 and the Nanotechnology Platform Program of the Ministry of Education, Culture, Sports, Science, and Technology, Japan.
\end{acknowledgements}

%\vspace{1cm}

\end{document}